\documentclass[twocolumn]{aastex631}

\usepackage{amssymb}
\usepackage{mathrsfs}
\usepackage{amsmath}

\begin{document}

\title{Phase-resolved Spectroscopy of Low-frequency Quasi-periodic Oscillations from the Newly Discovered Black Hole X-ray Binary Swift J1727.8--1613}

\correspondingauthor{Qing-Cang Shui}
\email{shuiqc@ihep.ac.cn}
\correspondingauthor{Shu Zhang}
\email{szhang@ihep.ac.cn}
\correspondingauthor{Jing-Qiang Peng}
\email{pengjq@ihep.ac.cn}

\author[0000-0001-5160-3344]{Qing-Cang Shui}
\affiliation{Key Laboratory of Particle Astrophysics, Institute of High Energy Physics, Chinese Academy of Sciences, 100049, Beijing, China}
\affiliation{University of Chinese Academy of Sciences, Chinese Academy of Sciences, 100049, Beijing, China}

\author{Shu Zhang}
\affiliation{Key Laboratory of Particle Astrophysics, Institute of High Energy Physics, Chinese Academy of Sciences, 100049, Beijing, China}

\author[0000-0002-5554-1088]{Jing-Qiang Peng}
\affiliation{Key Laboratory of Particle Astrophysics, Institute of High Energy Physics, Chinese Academy of Sciences, 100049, Beijing, China}
\affiliation{University of Chinese Academy of Sciences, Chinese Academy of Sciences, 100049, Beijing, China}

\author[0000-0001-5586-1017]{Shuang-Nan Zhang}
\affiliation{Key Laboratory of Particle Astrophysics, Institute of High Energy Physics, Chinese Academy of Sciences, 100049, Beijing, China}
\affiliation{University of Chinese Academy of Sciences, Chinese Academy of Sciences, 100049, Beijing, China}

\author[0000-0001-8768-3294]{Yu-Peng Chen}
\affiliation{Key Laboratory of Particle Astrophysics, Institute of High Energy Physics, Chinese Academy of Sciences, 100049, Beijing, China}

\author[0000-0001-9599-7285]{Long Ji}
\affiliation{School of Physics and Astronomy, Sun Yat-Sen University, Zhuhai, 519082, China}

\author[0000-0003-3188-9079]{Ling-Da Kong}
\affiliation{Institut f\"{u}r Astronomie und Astrophysik, Kepler Center for Astro and Particle Physics, Eberhard Karls, Universit\"{a}t, Sand 1, D-72076 T\"{u}bingen, Germany}

\author[0000-0001-7584-6236]{Hua Feng}
\affiliation{Key Laboratory of Particle Astrophysics, Institute of High Energy Physics, Chinese Academy of Sciences, 100049, Beijing, China}

\author{Zhuo-Li Yu}
\affiliation{Key Laboratory of Particle Astrophysics, Institute of High Energy Physics, Chinese Academy of Sciences, 100049, Beijing, China}


\author[0000-0002-6454-9540]{Peng-Ju Wang}
\affiliation{Institut f\"{u}r Astronomie und Astrophysik, Kepler Center for Astro and Particle Physics, Eberhard Karls, Universit\"{a}t, Sand 1, D-72076 T\"{u}bingen, Germany}

\author[0000-0003-4856-2275]{Zhi Chang}
\affiliation{Key Laboratory of Particle Astrophysics, Institute of High Energy Physics, Chinese Academy of Sciences, 100049, Beijing, China}


\author[0000-0002-0638-088X]{Hong-Xing Yin}
\affiliation{Shandong Key Laboratory of Optical Astronomy and Solar-Terrestrial Environment, School of Space Science and Physics, Institute of Space Sciences, Shandong University, Weihai, Shandong 264209, China}

\author[0000-0002-9796-2585]{Jin-Lu Qu}
\affiliation{Key Laboratory of Particle Astrophysics, Institute of High Energy Physics, Chinese Academy of Sciences, 100049, Beijing, China}

\author[0000-0002-2705-4338]{Lian Tao}
\affiliation{Key Laboratory of Particle Astrophysics, Institute of High Energy Physics, Chinese Academy of Sciences, 100049, Beijing, China}

\author[0000-0002-2749-6638]{Ming-Yu Ge}
\affiliation{Key Laboratory of Particle Astrophysics, Institute of High Energy Physics, Chinese Academy of Sciences, 100049, Beijing, China}


\author[0000-0003-4498-9925]{Liang Zhang}
\affiliation{Key Laboratory of Particle Astrophysics, Institute of High Energy Physics, Chinese Academy of Sciences, 100049, Beijing, China}





\author{Jian Li}
\affiliation{CAS Key Laboratory for Research in Galaxies and Cosmology, Department of Astronomy, University of Science and Technology of China, Hefei 230026, China}
\affiliation{School of Astronomy and Space Science, University of Science and Technology of China, Hefei 230026, China}

\begin{abstract}
Low-frequency quasi-periodic oscillations (LFQPOs) are commonly observed in X-ray light curves of black hole X-ray binaries (BHXRBs); however, their origin remains a topic of debate. In order to thoroughly investigate variations in spectral properties on the QPO timescale, we utilized the Hilbert-Huang transform technique to conduct phase-resolved spectroscopy across a broad energy band for LFQPOs in the newly discovered BHXRB Swift J1727.8--1613. This is achieved through quasi-simultaneous observations from Neutron star Interior Composition ExploreR (NICER), Nuclear Spectroscopic Telescope ARray (NuSTAR), and Hard X-ray Modulation Telescope (Insight-HXMT). Our analysis reveals that both the non-thermal and disk-blackbody components exhibit variations on the QPO timescale, with the former dominating the QPO variability. For the spectral parameters, we observe modulation of the disk temperature, spectral indices, and reflection fraction with the QPO phase with high statistical significance ($\gtrsim5\sigma$). Notably, the variation in the disk temperature is found to precede the variations in the non-thermal and disk fluxes by $\sim0.4-0.5$ QPO cycles. We suggest that these findings offer further evidence that the type-C QPO variability is a result of geometric effects of the accretion flow.

\end{abstract}



\section{Introduction} \label{sec:intro}
Black hole X-ray binaries (BHXRBs) are commonly observed to undergo outbursts, in which the source exhibit transitions in their X-ray spectral and timing variability properties \citep{2006ARA&A..44...49R,2007A&ARv..15....1D}. These outbursts can be categorized into distinct canonical states, each characterized by unique X-ray spectral and variability properties \citep[][]{2005A&A...440..207B,2005Ap&SS.300..107H}. In the hard state, the emitting spectrum is dominated by a nonthermal power-law component, whereas in the soft state, a thermal multi-temperature blackbody component becomes dominant \citep{2006ARA&A..44...49R}. The multi-temperature blackbody component is believed to originate from a geometrically thin and optically thick accretion disk \citep{1973A&A....24..337S,1974MNRAS.168..603L}. The nonthermal power-law component is produced by the Comptonization of soft photons in an extended cloud consisting of hot electrons, commonly referred to as the X-ray corona \citep{1980A&A....86..121S,1994ApJ...434..570T,1996MNRAS.283..193Z}, and/or in a jet base \citep{2005ApJ...635.1203M,2021NatCo..12.1025Y}. Additionally, reflection features are often detected in the X-ray spectrum as a reflection component, formed by a portion of Comptonized photons irradiating the disk and then being scattered into the line of sight. This component includes abundant features, like broad emission lines and Compton hump, etc. \citep{2005MNRAS.358..211R,2010MNRAS.409.1534D,2014ApJ...782...76G}. In the time domain, low frequency quasi-periodic oscillations \citep[LFQPOs, roughly 0.1--30 Hz,][]{1989ARA&A..27..517V} have been routinely observed in most BHXRBs. These features are characterized by a narrow peak with finite width in the power spectral density (PSD), with the classifications of type A, B, and C based on the centroid frequency, quality factor and root-mean-square (rms) amplitude \citep{1999ApJ...526L..33W,2005ApJ...629..403C,2006ARA&A..44...49R}.

In the past few decades, various theoretical models, based on either the geometric effect or the intrinsic variability of the accretion flow, have been proposed to explain the physical origin of LFQPOs. Intrinsic models include the trapped corrugation modes, the Accretion-ejection instability model (AEI), and the Two-Component Advection Flow model, among others \citep{1990PASJ...42...99K,1999PhR...311..259W,1999A&A...349.1003T,1996ApJ...457..805M}. Geometric models, on the other hand, are mainly related to the relativistic precession of the accretion flow due to the frame dragging effect, and are developed into the relativistic precession model \citep[RPM;][]{1999ApJ...524L..63S}, L-T precession of the hot flow \citep{2009MNRAS.397L.101I} and jet precession model \citep{2016MNRAS.460.2796S,2019MNRAS.485.3834D,2021NatAs...5...94M,2023ApJ...948..116M}. Observational studies have shown that the variability of type-C QPOs generally increases with photon energy \citep{2017ApJ...845..143Z,2018ApJ...866..122H,2020JHEAp..25...29K,2020MNRAS.494.1375Z} and no prominent disc-like component exists in the rms spectra \citep{2006MNRAS.370..405S,2014MNRAS.438..657A,2016MNRAS.458.1778A,2023ApJ...957...84S}. Additionally, \citet{2021NatAs...5...94M} reported the discovery of tpye-C QPOs above 200 keV, indicating a strong relationship between type-C QPOs and Comptonized emission. Recently, there has been some evidence suggesting a geometric origin of LFQPOs. With large samples of observations, the inclination dependence of amplitudes and time lags of type C and B QPOs has been determined \citep[see][]{2015MNRAS.447.2059M,2017MNRAS.464.2643V}. Furthermore, reflection variability revealed with phase-resolved spectroscopy provides further support for a geometrical origin of type-C QPOs \citep[see][]{2015MNRAS.446.3516I,2016MNRAS.461.1967I,2017MNRAS.464.2979I,2022MNRAS.511..255N,2023ApJ...957...84S}. We refer readers to \citet{2019NewAR..8501524I} for recent reviews of observations and theories of LFQPOs.

Swift J1727.8–1613, a new bright X-ray transient, was discovered on August 24, 2023, by the Burst Alert Telescope of Swift (Swift/BAT) and initially identified as GRB 230824A \citep{2023GCN.34537....1P}. This source underwent a giant outburst, and was also detected by Gas Slit Camera of the Monitor of All-sky X-ray Image (MAXI/GSC), with a peak flux exceeding 7 Crab in the MAXI/GSC 2--20 keV energy band \citep{2023ATel16205....1N,2023ATel16206....1N}. Subsequently, this outburst was extensively monitored at optical and radio bands \citep{2023ATel16208....1C,2023ATel16211....1M}. Based on the observations from Imaging X-ray Polarimeter Explorer (IXPE), a significant polarization was detected with a polarization degree (PD) of $4.71\%\pm0.27\%$ and polarization angle (PA) of $2^\circ.6\pm1^\circ.7$ in 2--8 keV energy range \citep{2023ATel16242....1D}. During this outburst, significant type-C QPOs were detected by multiple X-ray instruments \citep{2023ATel16235....1K,2023ATel16219....1D,2023arXiv231006697M,2023ApJ...958L..16V,2024MNRAS.529.4624Y,2024ApJ...968...76I}. In this Letter, we conduct a phase-resolved analysis of the type-C QPOs observed in Swift J1727.8--1613 quasi-simultaneously by NICER, NuSTAR and Insight-HXMT. We provide an overview of the observations and our data reduction in Section~\ref{sec:2}, followed by the presentation of the timing and spectral analyses in Section~\ref{sec:3}. Finally, we discuss and summarize these results in Section~\ref{sec:4}.

\section{Observations and Data Reduction} \label{sec:2}
\subsection{Observations}
\begin{figure*}
\centering
    \includegraphics[width=\textwidth]{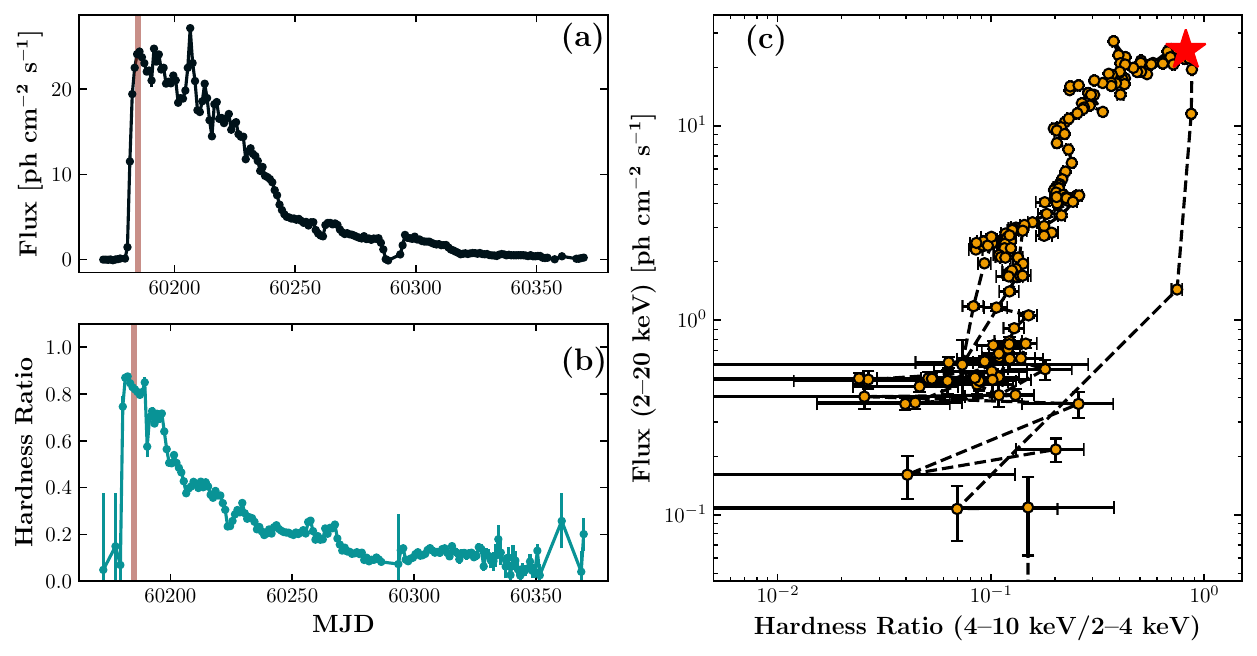}
    \caption{Evolution of X-ray properties of Swift J1727.8--1613 during the 2023--2024 outburst, measured by MAXI/GSC. The red vertical lines plotted in (a) and (b) represent the date of the quasi-simultaneous observations of NICER, NuSTAR and Insight-HXMT. (a) MAXI light curve in the 2--20 keV energy range. (b) Evolution of the hardness ratio of the X-ray count rate in the 4--10 keV energy band to that in the 2--4 keV energy band. (c) Hardness-intensity diagram (HID). The position of the source on the diagram during the quasi-simultaneous observations of NICER, NuSTAR and Insight-HXMT is marked with the red star.}
    \label{fig:1}
\end{figure*}

\begin{figure}
\centering
\includegraphics[width=\linewidth]{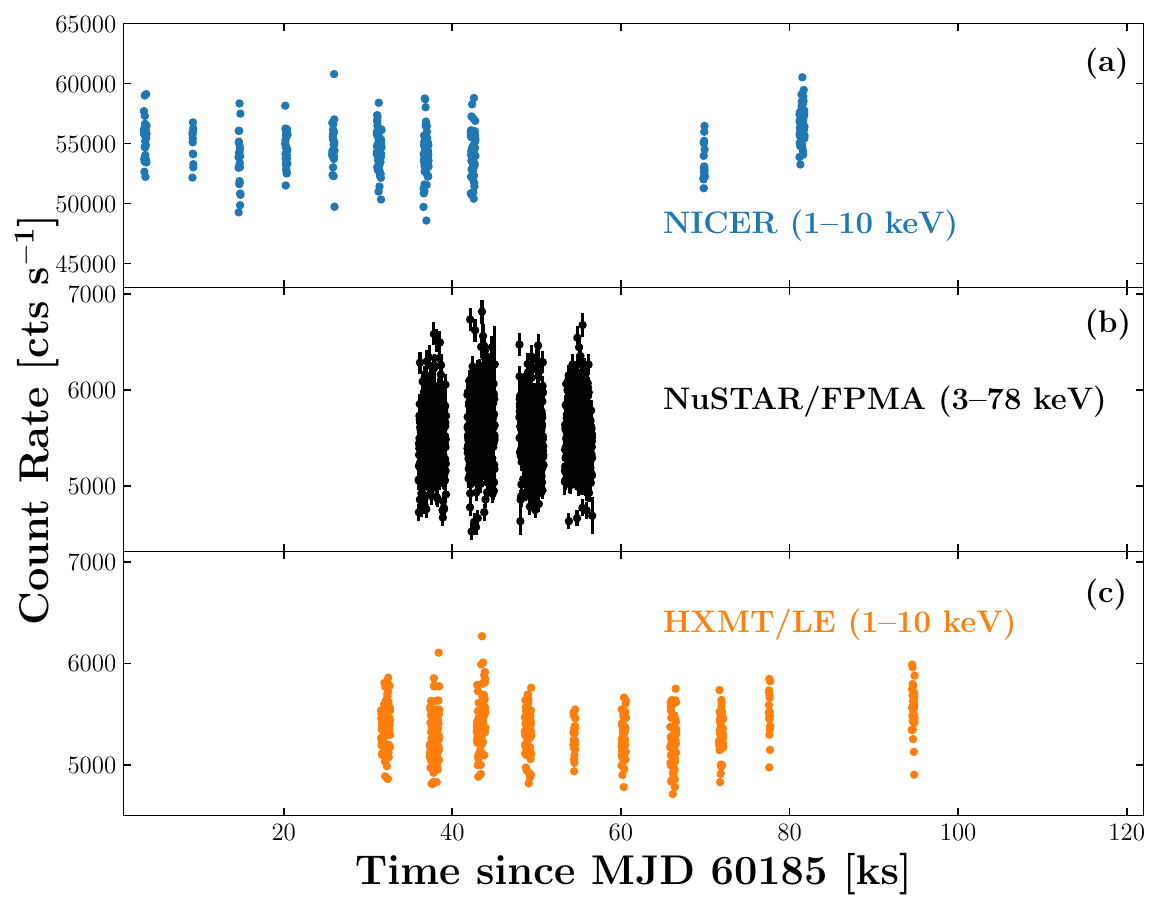}
    \caption{Light curves (counts per second) of Swift J1727.8--1613 with the 10-s time resolution from the quasi-simultaneous observations of NICER (1--10 keV), NuSTAR/FPMA (3--78 keV) and Insight-HXMT/LE (1--10 keV).}
    \label{fig:2}
\end{figure}

As mentioned in Section~\ref{sec:intro}, modeling the reflection component in phase-resolved analysis is crucial for diagnosing QPO models. Therefore, our study includes NuSTAR data to provide further constraints on the reflection component. NICER, NuSTAR and Insight-HXMT conducted quasi-simultaneous observations of the newly discovered BHXRB Swift J1727.8--1613 quasi-simultaneously on MJD 60185. In Figure~\ref{fig:1} (a) and (b), the timing of these observations is shown with respect to the source flux and hardness ratio variations as measured by MAXI/GSC. Figure~\ref{fig:1} (c) positions the observations on a hardness-intensity diagram (HID), indicating that the quasi-simultaneous observations were carried out at the beginning of the hard-to-soft state transition, where the luminosity of the source is close to the peak value. Figure~\ref{fig:2} displays the light curves of the NICER, NuSTAR/FPMA and Insight-HXMT/LE observations with a 10-s time resolution. Further details of these quasi-simultaneous observations can be found in Table~\ref{tab:1}.

\begin{table}[]
    \centering
    \caption{Log of NICER, Insight-HXMT, and NuSTAR Observations of Swift J1727.8--1613 Used in This Work.\label{tab:1}}
    \begin{tabular}{ccc}
    \hline
    \hline
    Instrument & Start Time & Exposure Time\\ 
    & (MJD) & (s)\\
    \hline
    NICER & 60185.04 & 3627 \\
    Insight-HXMT/LE & 60185.31 & 11450 \\
    Insight-HXMT/ME & 60185.31 & 8664 \\
    Insight-HXMT/HE & 60185.31 & 9875 \\
    NuSTAR/FPMA & 60185.42 & 923 \\
    NuSTAR/FPMB & 60185.42 & 1012 \\
    \hline
    \hline
    \end{tabular}
\end{table}

\subsection{Data Reduction}
\subsubsection{NICER}
NICER, a payload on the International Space Station (ISS), is dedicated to timing and spectroscopic analysis at soft X-rays \citep[0.2--12 keV;][]{2016SPIE.9905E..1HG}. During the 2023 outburst of Swift J1727.8--1613, NICER conducted frequent observations. In this study, we analyze an observation with ObsID 6203980105, which was carried out almost simultaneously with NuSTAR and Insight-HXMT observations. Since 2023 May 22, NICER has experienced a light leak issue that results in increased background noise during orbit day observations when solar light reaches the X-ray detectors. The regular detector resets, known as undershoots, which are characteristic of a functioning Focal Plane Module (FPM), occur more frequently due to the presence of solar light and subsequently lead to degraded spectral resolution and increased noise levels at lower energies\footnote{\url{https://heasarc.gsfc.nasa.gov/docs/nicer/analysis_threads/undershoot-intro/}} \citep[see also][]{2024ApJ...968...76I}. The data reduction is performed by using the official software \texttt{HEASOFT V6.33/NICERDAS} package with the latest CALDB \texttt{xti20221001}. We first reprocess the NICER data with the standard pipeline processing tool \texttt{nicerl2}. In this step, the default maximum undershoot rate criterion of 500 undershoots per second is used for the observation. Subsequently, the light curves and energy spectrum are extracted using \texttt{nicerl3-lc} and \texttt{nicerl3-spect} tools, respectively. The \texttt{nibackgen3C50} model \citep{2022AJ....163..130R} is used to estimate the background corresponding to the observation. Here, FPM 14 and 34 are always excluded, since they often exhibit episodes of increased detector noise. The adopted energy range in the spectral analysis is 1--10 keV, and a systematic error of 1 per cent is incorporated for 1--4 keV energy band to reduce the effects from calibration uncertainties at low energies.

\subsubsection{NuSTAR}
NuSTAR, the first mission to use focusing telescopes to image the sky in the high-energy X-ray (3--79 keV) region of the electromagnetic spectrum, was launched at 9 am PDT, June 13, 2012 \citep{2013ApJ...770..103H}. The data are processed using the NuSTAR Data Analysis Software (\texttt{NuSTARDAS v2.1.2}) with the latest calibration data base (\texttt{CALDB 20230613}). At first, the standard pipeline routine \texttt{nupipeline} is used for data reduction. Since Swift J1727.8--1613 was very bright, an additional status expression \texttt{STATUS == b0000xxx00xxxx000}\footnote{\url{https://heasarc.gsfc.nasa.gov/docs/nustar/analysis/}} is applied in this step. Then, \texttt{nuproducts} is used to generate light curves, spectra, response matrix, and ancillary response files, where source and background events are extracted from a 160 arcsec circular region and a 130 arcsec source-free circular region, respectively. Finally, we group the spectra via the \texttt{grppha} task to ensure a signal-to-noise ratio of at least five for each bin.

\subsubsection{Insight-HXMT}
Insight-HXMT is the first Chinese X-ray astronomy satellite \citep{2014SPIE.9144E..21Z,2020SCPMA..6349502Z}, launched on June 15, 2017. The science payload of Insight-HXMT allows for observations across a broad energy band (1--250 keV) using three telescopes: the High Energy X-ray telescope (HE, composed of NaI/CsI, 20--250 keV), the Medium Energy X-ray telescope (ME, with a Si pin detector, 5--30 keV), and the Low Energy X-ray telescope (LE, using an SCD detector, 0.7--13 keV). For additional information about the Insight-HXMT mission, please refer to \citet{2020SCPMA..6349502Z}, \citet{2020SCPMA..6349503L}, \citet{2020SCPMA..6349504C} and \citet{2020SCPMA..6349505C}. 
The event data are extracted by using Insight-HXMT \texttt{Data Analysis Software v2.05}, along with the current calibration model \texttt{v2.06}\footnote{\url{http://hxmtweb.ihep.ac.cn/software.jhtml}} and the standard Insight-HXMT \texttt{Data Reduction Guide v2.05}\footnote{\url{http://hxmtweb.ihep.ac.cn/SoftDoc.jhtml}}, and filtered with the following series of criteria recommended by the Insight-HXMT team: (1) the elevation angle (ELV) is greater than $10^{\circ}$; (2) the geometric cutoff rigidity (COR) is greater than 8 GeV; (3) the pointing position offset is less than $0.04^{\circ}$; and (4) the good time intervals (GTIs) are at least 300 s away from the South Atlantic Anomaly (SAA). The backgrounds are produced from blind detectors using the \texttt{LEBKGMAP}, \texttt{MEBKGMAP}, and \texttt{HEBKGMAP} tools, version 2.0.9, based on the standard Insight-HXMT background models \citep{2020JHEAp..27...14L,2020JHEAp..27...24L,2020JHEAp..27...44G}. The observations of Swift J1727.8--1613 were affected by temporary calibration issues in the LE data, impacting the spectral analysis at the soft energies \citep[see][]{2024ApJ...960L..17P}. As a result, this study only utilizes the LE data for timing analysis, and does not include spectral analysis with the LE data. Following the recommendation of the Insight-HXMT calibration group, the ME and HE spectra are rebinned as follows: (1) ME: Channels 0–1023 are grouped with a rebin factor of 2; (2) HE: Channels 0–255 are grouped with a rebin factor of 2. Additionally, considering the current accuracy of the instrument calibration, we add 0.5 per cent, and 1 per cent systematic errors to the energy spectra for ME and HE, respectively.

\section{Analysis and Results}
\label{sec:3}
\subsection{Timing Analysis}
\begin{figure*}
\centering
\begin{minipage}[c]{0.46\linewidth}
\centering
    \includegraphics[width=\linewidth]{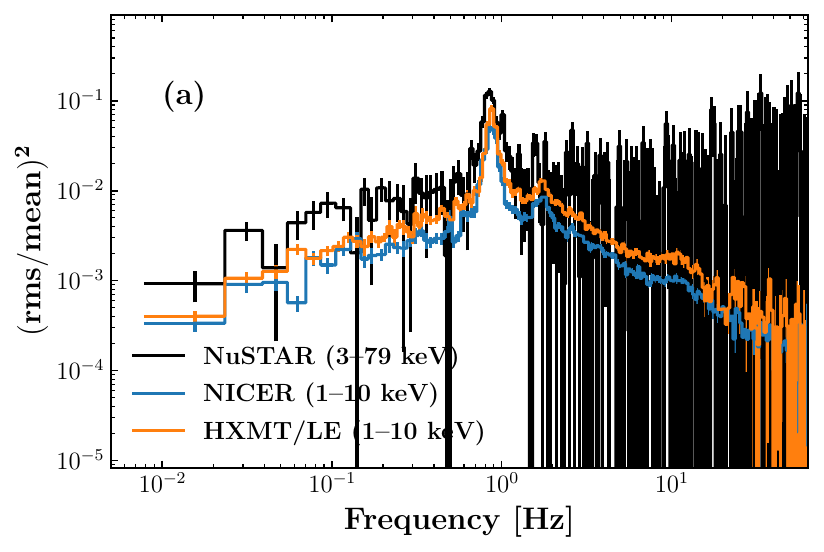}
\end{minipage}
\begin{minipage}[c]{0.45\linewidth}
\centering
    \includegraphics[width=\linewidth]{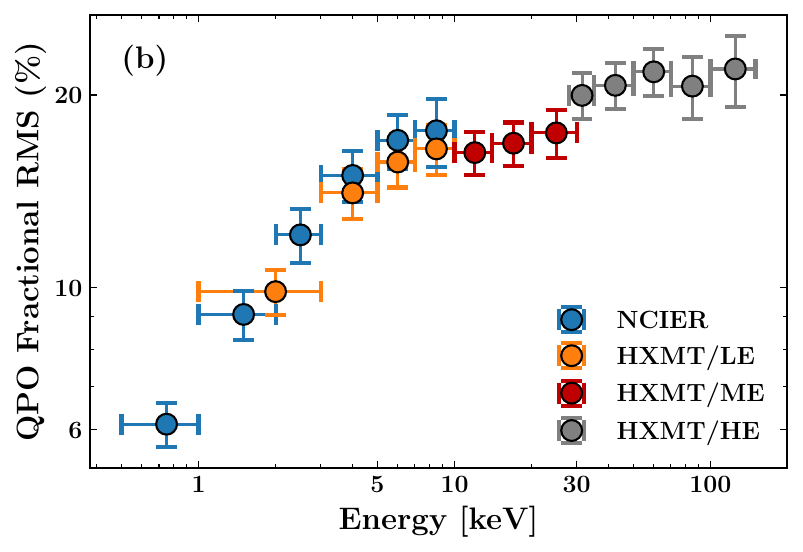}
\end{minipage}\\
\centering
    \begin{minipage}[c]{0.45\linewidth}
\centering
    \includegraphics[width=\linewidth]{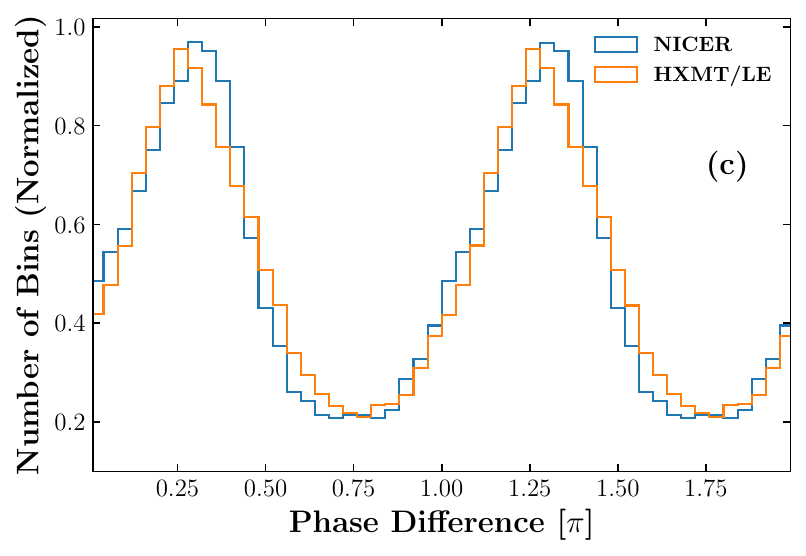}
\end{minipage}
\centering
    \begin{minipage}[c]{0.45\linewidth}
\centering
    \includegraphics[width=\linewidth]{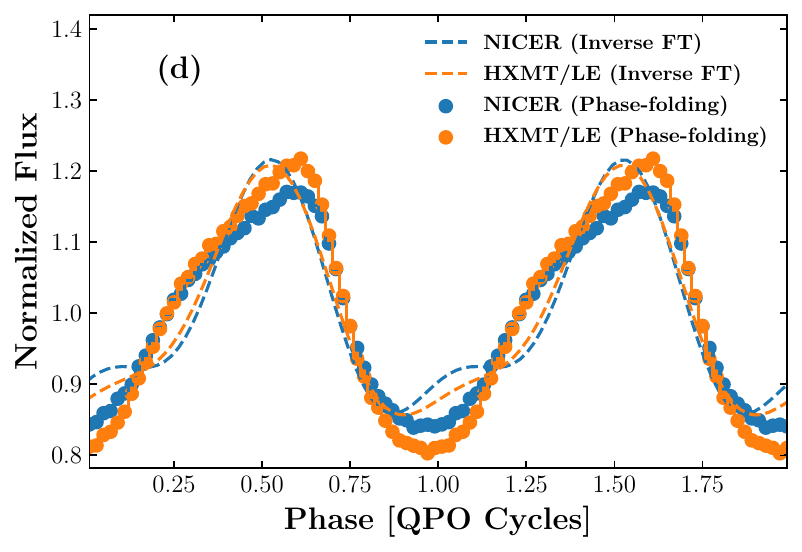}
\end{minipage}
    \caption{Panel (a): 1--10 keV power density spectrum for NICER (blue) and Insight-HXMT/LE (orange), and 3--79 keV co-spectrum between NuSTAR FPMA and FPMB (black). Panel (b): energy dependence of QPO fractional rms using NICER, HXMT LE, ME and HE data. Panel (c): histogram of measured values of the phase difference between QPO harmonics obtained using the HHT analysis for NICER (blue) and HXMT/LE (orange) data sets (see the text for details). Panel (d): reconstructed QPO waveform derived from the HHT phase-folding (scatters) and inverse FT (dashed lines) methods for NICER (blue) and HXMT/LE (orange) data sets.} \label{fig:3}
\end{figure*}

We first perform the timing analysis for these quasi-simultaneous observations. The light curves with the 1/128-s time bin from the NICER and Insight-HXMT observations are used to generate the PSDs, with a constant Poisson noise level subtracted. As for the NuSTAR observation, we instead calculate the co-spectrum between the two independent focal plane modules, FPMA and FPMB \citep{2015ApJ...800..109B}, due to the NuSTAR dead time of $\tau_{\rm d}\approx2.5$ ms, which introduces instrumental features on the Poisson noise in a PSD calculated in the standard way. The co-spectrum, which is the real part of the cross-spectrum, does not include any Poisson noise contribution. We then proceed to correct the co-spectrum for the suppression of variability caused by the NuSTAR dead time using the simple formula \citep{2015ApJ...800..109B}:
\begin{equation}
    \frac{\rm rms_{det}}{\rm rms_{in}}\approx\frac{1}{1+\tau_{\rm d}r_{\rm in}}=\frac{r_{\rm det}}{r_{\rm in}}
\end{equation}
where $r_{\rm det}$ and $r_{\rm in}$ are the detected and intrinsic count rate, respectively. For this observation, the ratio of detected to intrinsic variability is $\rm{rms_{det}/rms_{in}=7.38\times10^{-2}}$. As shown in Figure~\ref{fig:3} (a), all PSDs display QPOs with a strong fundamental and relatively weaker second harmonic. The QPO fundamental frequency is detected at $\sim0.88$ Hz, and the NuSTAR co-spectrum is in good agreement with the NICER and Insight-HXMT data. As the total energy range of the NICER and Insight-HXMT data covers the energy band of NuSTAR, and there are no dead time effects in NICER and Insight-HXMT data, we concentrate solely on the NICER and Insight-HXMT observations for the subsequent detailed timing analysis. We produce PSDs in different energy bands and fit them with a sum of Lorentzian functions. The fractional rms of QPO fundamental is calculated from the integral of the corresponding Lorentzian function. Figure~\ref{fig:3} (b) shows the QPO rms as a function of photon energy from both NICER and Insight-HXMT data. It is evident that the QPO amplitude increases with energy in the $\sim1-150$ keV energy range, with a hint of an additional enhancement seen above 30 keV. This phenomenon is very different from that the QPO rms is observed to decrease sightly with energy above $\sim1$ keV and up to 200 keV in the LHS of MAXI J1820+070 \citep{2021NatAs...5...94M,2023ApJ...948..116M}, but it is consistent with findings in other sources, such as MAXI J1535--571 \citep{2018ApJ...866..122H,2020JHEAp..25...29K}.

QPO phase-resolved spectroscopy seeks to reveal spectral variations on QPO timescales. This task is complicated by the `quasi-’ nature of the oscillation, which hinders more direct approaches such as folding on a period. To construct the QPO phase-resolved spectrum, we adopt the procedure outlined by \citet{2023ApJ...957...84S} that performing the Hilbert-Huang transform (HHT) analysis. The HHT method, initially introduced by \citet{1998RSPSA.454..903H} as an adaptive data analysis technique, provides a powerful tool for studying signals with non-stationary periodicity \citep{1998RSPSA.454..903H,2008RvGeo..46.2006H}. This approach comprises two main components: mode decomposition and Hilbert spectral analysis (HSA). The mode decomposition aims to decompose a time series into several intrinsic mode functions (IMFs), while the HSA allows obtaining both the frequency and phase functions for the desired IMFs, such as the QPO component \citep[e.g.][]{2014ApJ...788...31H,2020ApJ...900..116H,2023ApJ...957...84S,2024ApJ...965L...7S}. We perform the HHT analysis on the light curves in the 1--10 keV energy band from NICER and Insight-HXMT/LE observations, which is presented in Appendix~\ref{appendix1}. By employing the variational mode decomposition \citep[VMD,][]{6655981}, the intrinsic variability of the fundamental and the second harmonics of the 0.88-Hz QPO can be identified as the second and third IMFs, respectively. Based on the Hilbert transform, we can obtain the instantaneous phase functions of the QPO fundamental and harmonic components. The phase difference between QPO fundamental and second harmonics, as defined in Equation (3) of \citet{2015MNRAS.446.3516I}, influences the underlying waveform of the QPO. We employ the method proposed by \citet{2023ApJ...957...84S} to calculate the phase difference between the second and third IMFs for each time bin. Figure~\ref{fig:3} (c) presents the distribution of the phase difference measured for each time bin. It is evident that, for both the NICER and Insight-HXMT/LE data sets, the phase difference is clearly distributed around a mean value, indicating that phases of the second and third IMFs are corrected, and the results obtained from the NICER and Insight-HXMT/LE are consistent. To determine the mean value of the phase difference, we use the method described in \citet{2015MNRAS.446.3516I}. This calculation yields the mean phase difference of $\langle\psi\rangle/\pi=0.262\pm0.002$ and $\langle\psi\rangle/\pi=0.279\pm0.002$ for NICER and Insight-HXMT/LE observations, respectively. As proposed by \citet{2015MNRAS.446.3516I}, the mean value of the phase difference and the amplitudes of the QPO fundamental and harmonic components can be used to construct the underlying waveform through the inverse Fourier transform (FT). Additionally, the HHT phase folding method can also give the QPO waveform. The waveforms obtained from both the phase folding and inverse FT methods are presented in Figure~\ref{fig:3} (d). It is worth noting that, in comparison to the Fourier method, the HHT phase-folding method presents both advantages and disadvantages. For example, as mentioned by \citet{2022hxga.book..113I} and \citet{2023ApJ...957...84S}, the contribution of the harmonic component in the waveform derived from the HHT phase folding method is suppressed. Nevertheless, it is important to note that while the current Fourier method is limited to introducing only the QPO second harmonic, the HHT phase-folding method has the capability to incorporate higher-order harmonics (i.e., greater than twice the QPO frequency), albeit with some degree of suppression. For the analysed QPO signals in our study, waveforms obtained from the two methods are consistent, displaying a non-sinusoidal nature, characterized by a \textit{slow rise and fast decay} feature.

\subsection{Energy Spectral Analysis}
\begin{table}[]
\begin{center}
    \caption{Phase-averaged spectral fitting results for Swift J1727.8--1613 on MJD 60185.\label{tab:2}}
    \begin{tabular}{lccc}
    \hline
    \hline
    Model & Parameter & NICER$^{\rm a}$ & Joint View$^{\rm b}$\\ 
    \hline
    TBABS & $N_{\rm H}$ ($10^{21} {\rm cm^{-2}}$) & $2.08^{+0.02}_{-0.01}$ & $1.84^{+0.01}_{-0.01}$\\
    DISKBB & $T_{\rm in}$ (keV) & $0.41^{+0.01}_{-0.01}$ & $0.46^{+0.01}_{-0.01}$\\
          & $N_{\rm disk}^{\rm c}$ ($10^4$) & $9.78^{+2.06}_{-1.13}$ & $5.81^{+0.42}_{-0.62}$ \\
    \hline
    RELXILL & $i$ ($^\circ$) & $36.81^{+3.25}_{-2.34}$ & $37.39^{+1.79}_{-1.21}$\\
            & $a*$ &  $0.92^{+0.06}_{-0.09}$ & $0.97^{+0.02}_{-0.05}$\\
            & $q^{\rm d}$ & 3 (f) & 3 (f)\\
            & $R_{\rm in}$ ($R_{\rm ISCO}$)&1 (f)& 1 (f)\\
            & $R_{\rm out}$ ($R_{\rm g}$) &  1000 (f)    & 1000 (f)\\
            & $\Gamma_1$ & $1.49^{+0.01}_{-0.01}$ & $1.77^{+0.02}_{-0.02}$ \\
            & $R_{\rm f}$ & $0.08^{+0.02}_{-0.01}$ & $0.24^{+0.04}_{-0.03}$\\
            & $E_{\rm cut1}$ (keV) &30 (f) & $56.9^{+2.6}_{-2.8}$\\
            & $A_{\rm Fe}$ & 10 (f) & $3.55^{+0.51}_{-1.31}$\\
            & $\log_{10}(\xi)^{\rm e}$&  $3.12^{+0.16}_{-0.24}$    & $3.24^{+0.06}_{-1.31}$\\
            & $N_{\rm rel}^{\rm f}$&   $0.59^{+0.01}_{-0.01}$   & $0.31^{+0.01}_{-0.01}$\\
    \hline
    CUTOFFPL & $\Gamma_2$& -- & $0.98^{+0.02}_{-0.02}$\\ 
             & $E_{\rm cut2}$ (keV) & -- & $11.5^{+0.2}_{-0.3}$\\
             & $N_{\rm cut}^{\rm j}$ & -- & $9.47^{+0.44}_{-0.43}$\\
    \hline
    CRABCORR & $\Delta\Gamma_{\rm NICER}$  &  -- & 0 (f)\\
             & $N_{\rm NICER}$  & --  & 1 (f)\\
             & $\Delta\Gamma_{\rm ME}$  & --  & $0.025^{+0.006}_{-0.006}$\\
             & $N_{\rm ME}$  &  -- & $0.902^{+0.013}_{-0.012}$\\
             & $\Delta\Gamma_{\rm HE}$  & --& $0.021^{+0.029}_{-0.016}$\\
             & $N_{\rm HE}$  &  -- & $0.859^{+0.092}_{-0.048}$\\
             & $\Delta\Gamma_{\rm FPMA}$  & -- & $0.117^{+0.004}_{-0.004}$\\
             & $N_{\rm FPMA}$  &  -- & $1.218^{+0.008}_{-0.007}$\\
             & $\Delta\Gamma_{\rm FPMB}$  &  -- & $0.102^{+0.004}_{-0.004}$\\
             & $N_{\rm FPMB}$  & --  & $1.121^{+0.007}_{-0.008}$\\
    \hline
             &  $\chi^2/{\rm d.o.f}$ &  164.13/162 & 2882.38/2742\\
    \hline
    \hline
    \end{tabular}
\end{center}
\tablecomments{$^{\rm a}$ Spectral fitting results with NICER data with Model 2. $^{\rm b}$ Joint spectral fitting results with NICER, NuSTAR and Insight-HXMT data with Model 4. $^{\rm c}$ Normalization of \texttt{diskbb} model. $^{\rm d}$ The power-law index of the emissivity profile ($\epsilon\propto r^{-q}$). $^{\rm e}$ Log of the ionization parameter ($\xi$) of the accretion disc, where $\xi=L/nR^2$, with $L$ as the ionizing luminosity, $n$ as the gas density, and $R$ as the distance to the ionizing source. $^{\rm f}$ Normalization of \texttt{relxill} model. $^{\rm j}$ Normalization of \texttt{cutoffpl} model.}
\end{table}

\begin{figure*}
\centering
\begin{minipage}[c]{0.48\linewidth}
\centering
    \includegraphics[width=\linewidth]{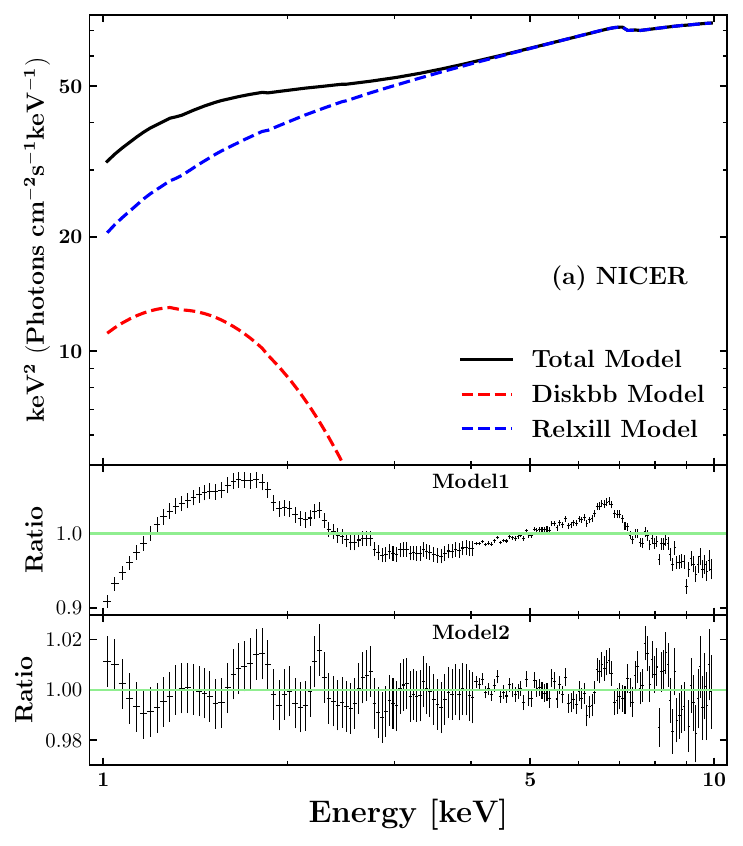}
\end{minipage}
\begin{minipage}[c]{0.48\linewidth}
\centering
    \includegraphics[width=\linewidth]{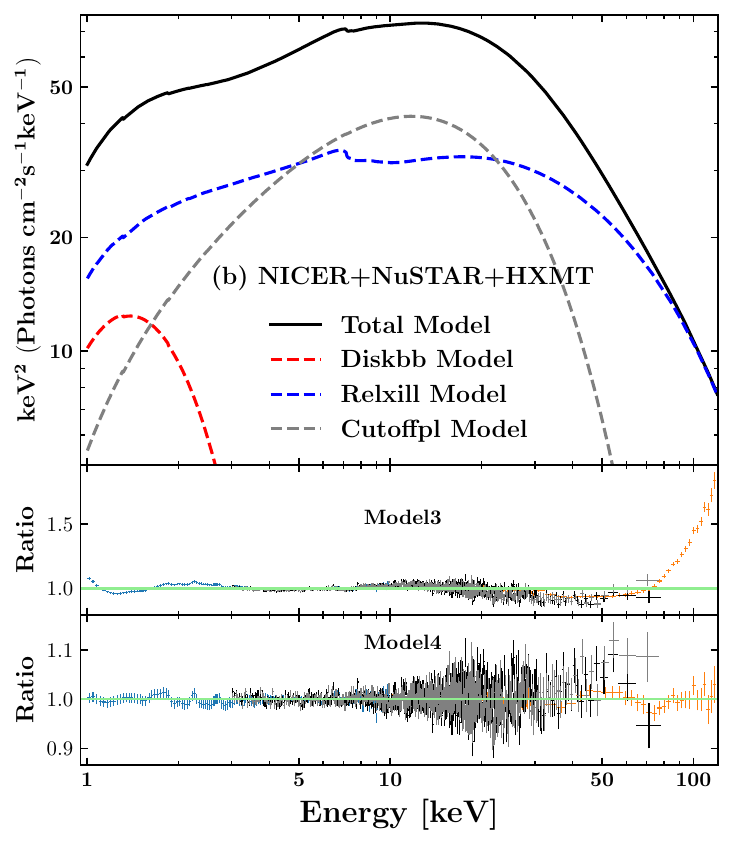}
\end{minipage}
    \caption{The best-fit models (top rows) and data-to-model ratios (lower rows) for different models. Panel (a) is for spectral analysis of NICER observation in 1--10 keV energy range, while panel (b) is for joint spectral analysis of NICER, NuSTAR and Insight-HXMT observations in 1--120 keV energy range.} \label{fig:4}
\end{figure*}

The spectral analysis is performed using \texttt{XPSEC v12.12.0} software package \citep{1996ASPC..101...17A}.  We first fit the phase-averaged energy spectrum from the NICER observation. The initial spectral model is \texttt{tbabs*powerlaw} (Model 1), where \texttt{tbabs} accounts for the galactic absorption, and \texttt{powerlaw} is a phenomenological power-law model. Upon fitting with Model 1, a broad iron line around 7 keV and a hump below 3 keV are clearly evident in the resulting data-to-model ratios (see Figure~\ref{fig:4}). Consequently, we use Model 2: \texttt{tbabs*(diskbb+relxill)}, to fit the energy spectrum, where \texttt{diskbb} is a nonrelativistic disk-blackbody component, and \texttt{relixll} is a reflection model from the \texttt{relxill} family\footnote{\url{https://github.com/thdauser/relxill/}} \citep{2014ApJ...782...76G,2014MNRAS.444L.100D}. In the reflection model \texttt{relxill}, we assume the canonical power-law emissivity profile, $\epsilon\propto r^{-3}$, for the thin disk. Since the inner and outer radius of the disk ($R_{\rm in}$ and $R_{\rm out}$), iron abundance ($A_{\rm Fe}$, in solar units) and $E_{\rm cut}$ cannot be well constrained only with NICER data, we fix these parameters at the innermost stable circular orbit, 1000 $R_{\rm g}$ ($R_{\rm g}=GM/c^2$), 10 and 30 keV, respectively. Fitting the NICER energy spectrum with Model 2, we obtain a $\chi^2/{\rm d.o.f}$ of 164.13/162. However, Figure~\ref{fig:4} (a) shows that the residuals are still strongly correlated below 4 keV. Specifically, it is evident that the correlated residual features are a edge-like shape near $\sim$2.2 keV and a Gaussian-like emission around 1.8 keV. These energies correspond to specific features in NICER's effective area versus energy, with the 1.8 keV and 2.2 keV features attributed to silicon and gold, respectively \citep[see also][]{2020ApJ...899...44W}. Therefore, we posit that these remaining features stem from systematic effects within NICER's calibration. Introduction of smaller systematic errors (e.g., 0.5\%) leads to a higher reduced chi-square value without significantly changes in the best-fitted parameter values. Subsequently, we conduct a joint spectral fitting of NICER, NuSTAR and Insight-HXMT data with Model 3: \texttt{tbabs*crabcorr*(diskbb+relxill)}. The cross-calibration among NICER, NuSTAR and Insight-HXMT is carried out by the model \texttt{crabcorr} \citep{2010ApJ...718L.117S,2020ApJ...899...44W,2022ApJ...928...11Z,2024ApJ...960L..17P,2024MNRAS.527.8029Y}, which applies corrections to both the slope of the power-law via the parameter $\Delta\Gamma$ and normalization, $N$, by multiplying each model spectrum by a power-law. In this fitting, we set $N$ and $\Delta\Gamma$ at fixed values of 1 and 0, respectively, for the NICER spectrum. However, jointly fitting the NICER, NuSTAR and Insight-HXMT spectra with Model 3 results in a poor goodness of fit ($\chi^2/{\rm d.o.f}=4021.38/2745$) and shows significant residual features in the high-energy band (see Figure~\ref{fig:4} (b)). To address this, we follow \citet{2024ApJ...960L..17P} to add a \texttt{cutoffpl} component to the fitting model (Model 4): \texttt{tbabs*crabcorr*(diskbb+relxill+cutoffpl)}. This model modification leads to a significant reduction in $\chi^2$ by 1139 with three extra free parameters, resulting in $\chi^2/{\rm d.o.f}=2882.38/2742$. Furthermore, the data-to-model ratios revealed no prominent residual features (see Figure~\ref{fig:4} (b)). As discussed in \citet{2024ApJ...960L..17P}, the origin of the additional \texttt{cutoffpl} component remains elusive, as it can be used to account for both an excess at $\sim$10--20 keV (as demonstrated in this analysis) and a hard tail above $\sim70$ keV. This suggests a potential degeneracy between the extra \texttt{cutoffpl} component and the incident continuum component of \texttt{relxill}. If this \texttt{cutoffpl} component does indeed indicate an additional physical component and the reflected emission truly originates from an individual component, different hybrid jet/corona configurations could qualitatively explain why the \texttt{cutoffpl} component is not being efficiently reflected \citep[see Figure 7 of][]{2024ApJ...960L..17P}. However, it is also possible that the reflector is illuminated by both components, but current spectroscopy is unable to distinguish between the reflected spectra from two distinct illuminators. Additionally, the extra \texttt{cutoffpl} component could potentially mimic the influence of a non-thermal population of electrons which makes the Comptonised spectrum distinct from a standard one. Additional details about the phase-averaged spectral fitting with Model 4 can be found in \citet{2024ApJ...960L..17P}.After conducting the aforementioned joint spectral analysis, we fix $E_{\rm cut}$ and $A_{\rm Fe}$ in the independent NICER spectral fitting to the best-fitting value from the joint spectral fitting, and find no significant changes in the obtained reduced chi-square and other free parameter values. Table~\ref{tab:2} summarizes results of independent spectral fitting of the NICER observation with Model 2 and joint spectral fitting with Model 4. The uncertainties are reported at the 90\% confidence level by employing the Markov Chain Monte Carlo (MCMC) method using the Goodman-Weare algorithm with 32 walkers and a total length of 50,000 \citep{2010CAMCS...5...65G}, and the initial 2000 elements are discarded as a burn-in period. 

\begin{figure*}
\centering
\begin{minipage}[c]{0.32\linewidth}
    \centering
    \includegraphics[width=\textwidth]{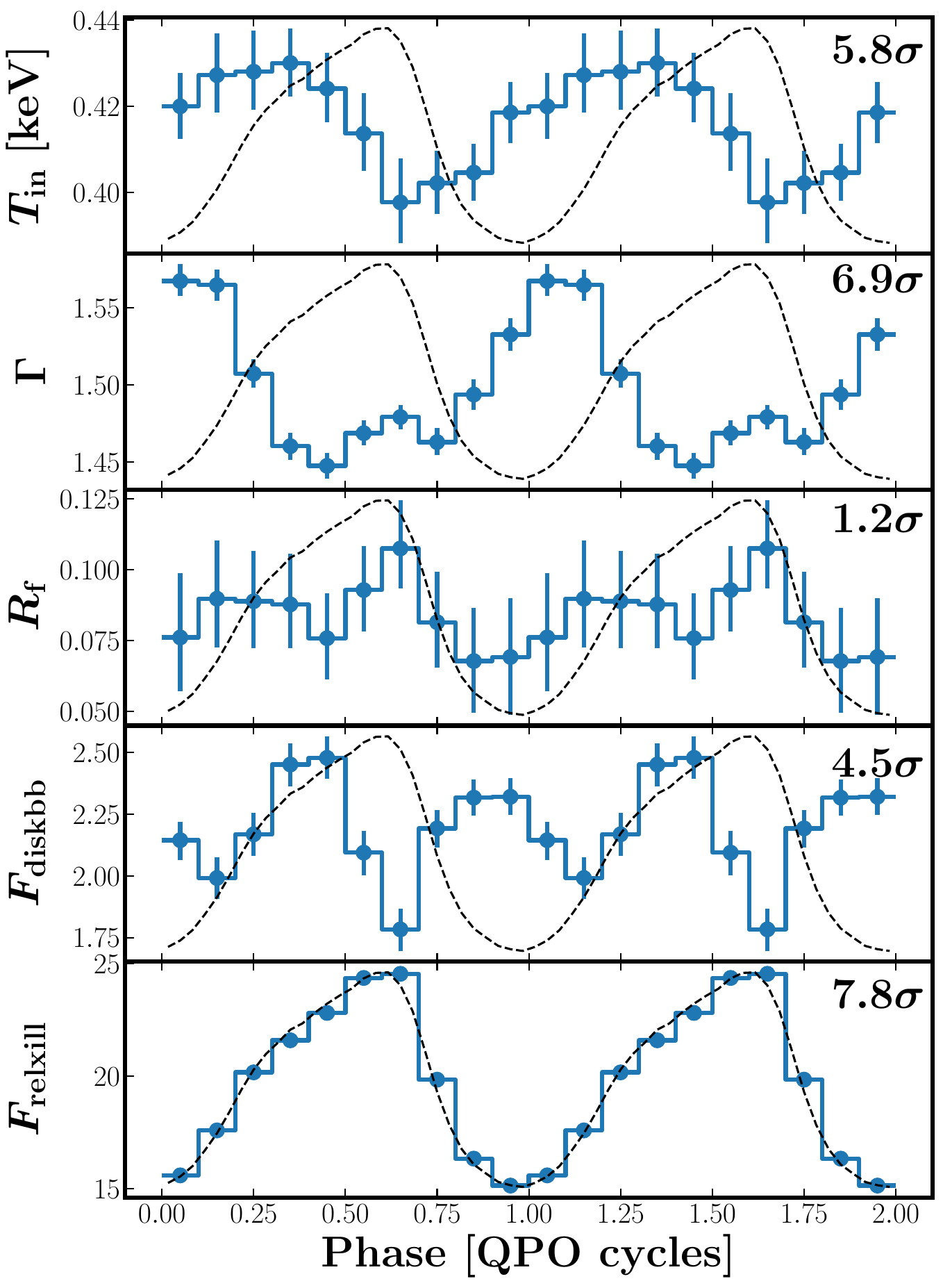}
\end{minipage}
\begin{minipage}[c]{0.32\linewidth}
    \centering
    \includegraphics[width=\textwidth]{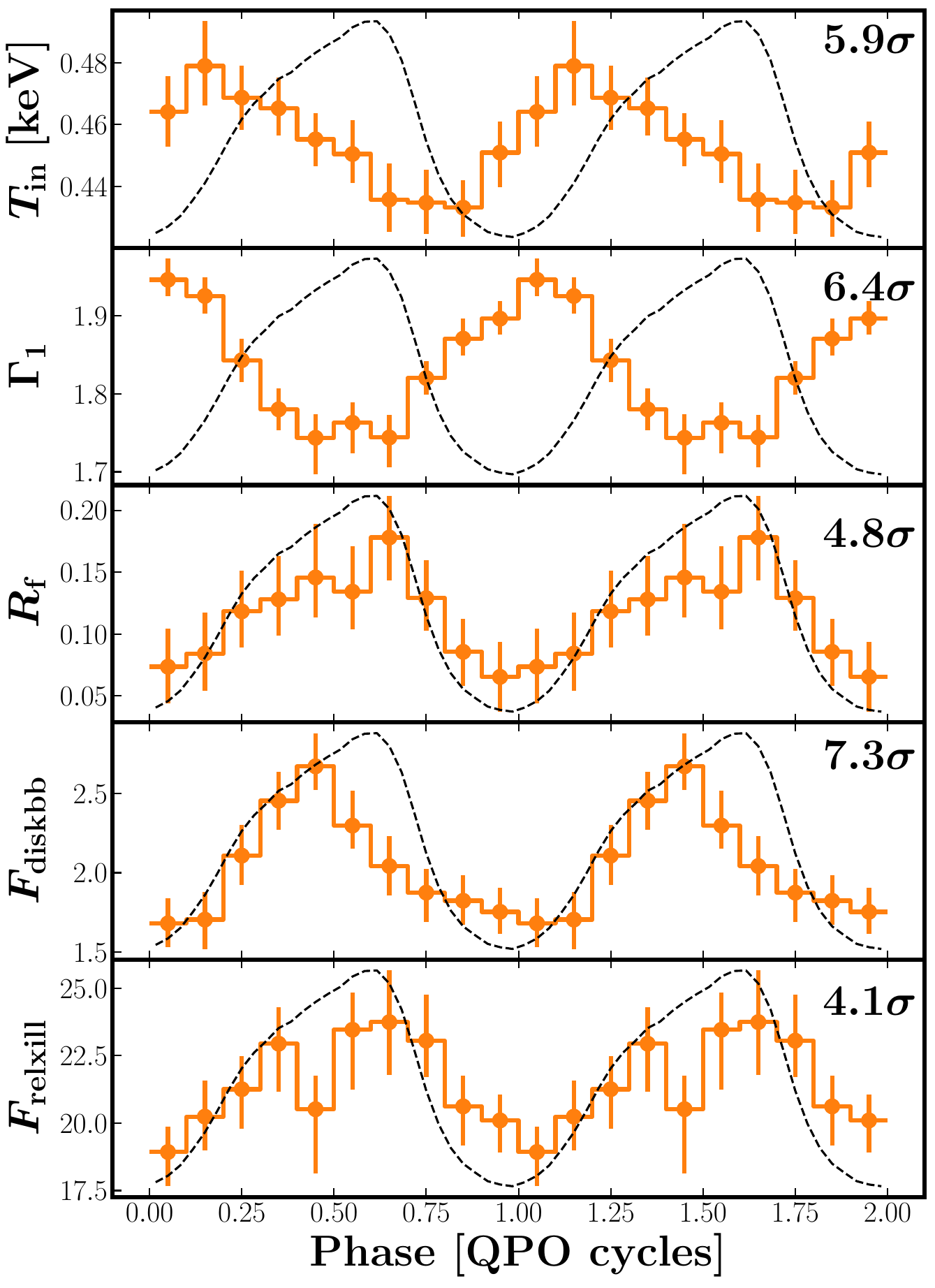}
\end{minipage}
\begin{minipage}[c]{0.32\linewidth}
    \centering
    \includegraphics[width=\textwidth]{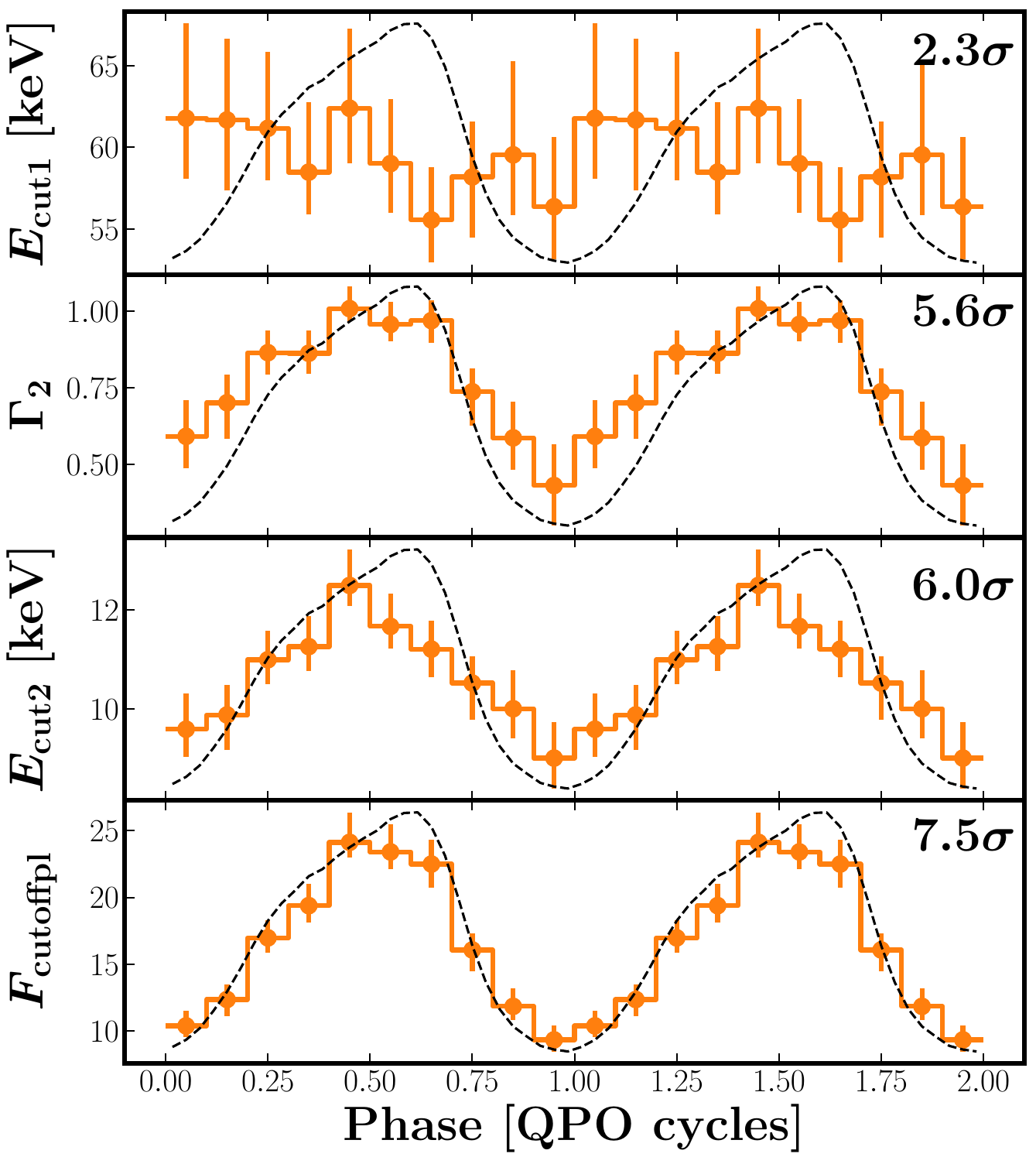}
\end{minipage}
    \caption{Phase dependence of the spectral parameters and model fluxes ($F_{\rm diskbb}$, $F_{\rm relxill}$ and $F_{\rm cutoffpl}$) from the phase-resolved spectral fitting of the NICER spectra (left panels) and joint NICER and Insight-HXMT spectra (middle and right panels). The shown model fluxes are in units of $10^{-8}\rm erg\ s^{-1}cm^{-2}$. Additionally, we include the phase-folding QPO waveform in each panel as the black dashed line. The statistical significance of each modulation is quoted in the corresponding plot.} \label{fig:5}
\end{figure*}

With the well-defined QPO phase given by the HHT analysis, we generate new GTIs corresponding to ten distinct phase intervals for NICER and Insight-HXMT observations, respectively. These new GTIs enable us extract spectra for each phase bin to perform subsequent phase-resolved spectral analysis. Due to the potential impact of NuSTAR dead time on the HHT analysis, we exclude the NuSTAR data from the phase-resolved spectral analysis. We perform independent spectral fitting with NICER observation with Model 2 and joint spectral fitting with NICER and Insight-HXMT observations with Model 4, respectively. In the phase-resolved spectral fittings, we fix the equivalent hydrogen column ($N_{\rm H}$), black hole spin parameter ($a_*$), disk inclination ($i$) and ionization parameter ($\log\xi$) to the best-fit values obtained from the phase-averaged spectral fittings. It is important to note that fixing the ionization parameter ($\xi$) while the illuminating flux changes is not entirely self-consistent, as $\xi$ should in principle increase when the reflected flux increases \citep[see e.g.][]{2014ApJ...782...76G,2019MNRAS.488..324I,2022MNRAS.511..255N}. However, we find that the ionization parameter turns out not sensitive to the phase-resolved spectral fitting and therefore is frozen at its best-fitted value in the phase-averaged spectral fitting. For the joint phase-resolved spectral fitting, we additionally fix the parameters ($\Delta\Gamma$ and $N$) from the \texttt{crabcorr} model to the best-fit values obtained from the joint phase-averaged spectral fittings. Furthermore, we employ \texttt{cflux} model in \texttt{XSPEC} to calculate unabsorbed flux contribution from each component. Similar to the phase-averaged spectral analysis, we utilize the MCMC method to compute the uncertainties. We find that the autocorrelation length in the MCMC chain typically consists of 1000 elements, resulting in the net number of independent samples in the parameter space being on the order of $10^4$. To further assess the convergence of the MCMC chain, we compare the one- and two-dimensional projections of the posterior distributions for each parameter from the first and second halves of the chain, and find no significant differences. In Appendix~\ref{appendix2}, we provide the contour maps and probability distributions for each free parameter in the joint phase-resolved spectral analysis.

Figure~\ref{fig:5} illustrates the phase dependence of the spectral parameters from the phase-resolved spectral fittings of NICER spectra and joint NICER and Insight-HXMT spectra. In these plots, we additionally include the QPO waveform in each panel as the black dashed line. We employ an F-test for each of these parameters to compare a fit with a constant model against the best-fitting model, in order to evaluate its variation with QPO phase and determine its statistical significance \citep[see also][]{2015MNRAS.446.3516I}. In the top-right corner of each panel in Figure~\ref{fig:5}, we provide the statistical significance with which each spectral parameter varies with QPO phase. It is evident that both phase-resolved spectral analyses reveal that $T_{\rm in}$ exhibits modulations with the QPO phase ($>5\sigma$). Furthermore, both results show substantial phase differences between the $T_{\rm in}$ and flux modulations. Since the joint spectral analysis is performed within a much boarder energy band (1--120 keV), we focus on the results from it. We observe strong modulations ($>5\sigma$) in the photon indices of \texttt{relxill} ($\Gamma_1$) and \texttt{cutoffpl} ($\Gamma_2$), but these two parameters exhibit opposite variation trends. The former lags the count rate modulation by $\sim0.5$ QPO cycles, while the later shows a same modulation phase as the count rate. Furthermore, phase modulations ($\sim5\sigma$) are also observed in reflection fraction ($R_{\rm f}$) and cut-off energy of \texttt{cutoffpl} ($E_{\rm cut2}$). Regarding the model fluxes, we find that the fluxes of \texttt{diskbb} and \texttt{cutoffpl} exhibit significant modulations ($>7\sigma$), with the later showing the largest absolute variability amplitude of $\sim 7\times10^{-8}\ {\rm erg\ cm^{-2}s^{-1}}$. Interestingly, a significant phase difference ($\sim0.3-0.4$ QPO cycles) between modulations of $T_{\rm in}$ and the \texttt{diskbb} flux ($F_{\rm disk}$) is observed. This indicates that $T_{\rm in}$ and $F_{\rm disk}$ do not satisfy a positive correlation over the QPO cycle.

\section{Discussion and Conclusion} \label{sec:4}
In this study, we have performed broad-band phase-resolved spectroscopy of type-C QPOs in the newly discovered BHXRB Swift J1727.8--1613, utilizing quasi-simultaneous NICER and Insight-HXMT observations. Additionally, to improve the constraints on the phase-averaged spectrum, we have included NuSTAR data in the phase-averaged spectral analysis. The results show that over the QPO period, the energy spectrum undergoes changes not only in normalization, but also in spectral shape. Specifically, parameters such as the temperature of the inner radius of the thin disk ($T_{\rm in}$), spectral indices ($\Gamma_1$ and $\Gamma_2$) and reflection fraction ($R_{\rm f}$) exhibit significant modulations. Notably, while the non-thermal component dominates the QPO variability, the disk-blackbody component also displays changes on the QPO timescale. Furthermore, a significant phase difference ($\sim0.3-0.4$ QPO cycles) between modulations of $T_{\rm in}$ and the flux of disk-blackbody component is observed. 

Figures~\ref{fig:5} demonstrates that the non-thermal component dominates the QPO variability. These results are consistent with the previous findings that the power-law component dominates the type-C QPO emission in BHXRBs \citep[e.g.][]{2006MNRAS.370..405S,2014MNRAS.438..657A,2016MNRAS.458.1778A}. Given that the nonthermal emission arises from the Compton upscattering of disk photons in a hot corona/jet base region \citep{1980A&A....86..121S,1994ApJ...434..570T,1999MNRAS.303L..11Z,2005ApJ...635.1203M}, our findings support models such as the corona's L-T precession model \citep{2009MNRAS.397L.101I}, precession of the jet base \citep{2021NatAs...5...94M} and the \texttt{vkompth} model \citep{2020MNRAS.492.1399K,2022MNRAS.515.2099B}. Phase-resolved analysis of the joint NICER and Insight-HXMT spectra reveals a significant modulation ($\sim5\sigma$) in the reflection fraction, which represents the ratio of Comptonized emission intensity illuminating the disk to the intensity directly reaching the observer \citep{2016A&A...590A..76D}. The observed variation in $R_{\rm f}$ suggests changes in accretion geometry over the cycle, providing strong evidence for the geometric origin of the QPO. Additionally, as the non-spherical corona/jet base precesses, the optical depth of the corona with respect to the observer ($\tau$) could vary over the precession period, leading to modulations in the photon index of the spectrum \citep{2023ApJ...957...84S}. Figure~\ref{fig:5} also shows strong modulations in both $\Gamma_1$ and $\Gamma_2$, with the two parameters exhibiting opposite variation trends. \citet{2024ApJ...960L..17P} proposed that the presence of an additional hard component indicates a joint jet-corona scenario in Swift J1727.8--1613, where the jet and corona could be somehow coupled and precess around the BH spin axis together \citep[see e.g.][]{2018MNRAS.474L..81L,2024ApJ...965L...7S}. The flux variation from the corona is attributed to the geometric wobble of the corona, altering the projection area with respect to the observer and modulating the observed X-ray flux \citep{2009MNRAS.397L.101I,2018ApJ...858...82Y,2023ApJ...943..165S}. In the case of jet precession, the flux variation is due to the Doppler boosting effects \citep{2021NatAs...5...94M}. In a hybrid corona/jet configuration to account for the nonthermal emission, a jet relevant to $\Gamma_1$ is set on top of a radially extended corona that corresponds to $\Gamma_2$ \citep[see][]{2024ApJ...960L..17P}. At the trough phases of the flux modulation, the corona precesses to an edge-on case and the harder jet base can be blocked by the surrounding corona, which can result in a softer spectral shape of the jet emission (i.e. larger $\Gamma_1$). On the other hand, a smaller $\Gamma_1$ is expected at the peak phases of the flux modulation when the corona is viewed face-on and the harder jet base becomes visible again. As suggested in \citet{2023ApJ...957...84S}, modulations in $\Gamma_2$ may be relevant to the changes of the optical depth of the corona: a harder spectrum shall be observed in an edge-on corona with a larger optical depth. Accordingly, although the fluxes from the corona and jet regions could vary synchronously, the modulations of the photon indices of the two spectral components can have different trends. 

\begin{figure}
\centering
    \includegraphics[width=\linewidth]{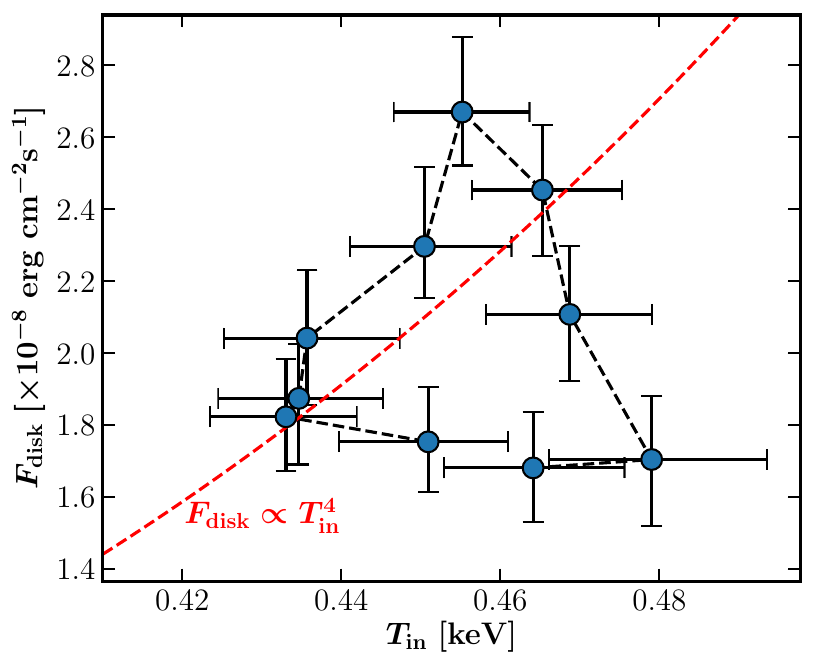}
    \caption{The disk flux vs. disk temperature $T_{\rm in}$ over the QPO cycle from the joint phase-resolved spectral fitting with Model 4. The dashed red line represents the relation of $F_{\rm disk}\propto T^4_{\rm in}$, which is predicted by the standard Shakura-Sunyaev disk model with a stable inner radius.}
    \label{fig:6}
\end{figure}

In terms of the variations in the disk component, both $T_{\rm in}$ and the disk flux show changes over the QPO cycle. We observe that $T_{\rm in}$ precedes the varying total flux by $\sim0.5$ QPO cycles. And more importantly, we find a significant phase difference ($\sim0.3-0.5$ QPO cycles) between modulations of $T_{\rm in}$ and the flux of disk-blackbody component. In Figure~\ref{fig:6}, we plot the disk flux versus $T_{\rm in}$ over the QPO cycle from the joint phase-resolved spectral fitting. This plot indicates that, over the QPO cycle, the disk emission completely deviates from the behavior of $F_{\rm disk}\tilde{\propto} T^4_{\rm in}$, which is predicted by the standard Shakura-Sunyaev disk with a stable inner radius. The inner radius of the disk, $R_{\rm in}$, typically changes on a viscosity timescale, $t_{\rm visc}$, which is given by
\begin{align}
    t_{\rm vsic}&\sim\frac{1}{\alpha\Omega}\left(\frac{R}{H}\right)^2\notag\\
    &\approx50{\rm s}\left(\frac{M}{10M_{\odot}}\right)\left(\frac{R}{R_{\rm g}}\right)^{3/2}\left(\frac{\alpha}{0.01}\right)^{-1}\left(\frac{H/R}{0.01}\right)^{-2},
\end{align}
where $\alpha$ is the viscosity parameter, $\Omega$ is the Kepler angular velocity, $R$ is the radius in the disk, $H$ is the vertical scale height of the disk and $M$ is the mass of the black hole \citep[see e.g.][]{2002apa..book.....F}. Considering that the parameter combination including a black hole mass of $10M_{\odot}$, $\alpha=0.01$, $H/R=0.01$ and $R=(10-20)R_{\rm g}$, the viscosity timescale is expected to be $\sim10^2-10^3$s, which is much longer than the QPO period ($\sim1$ s). This suggests that the inner radius of the disk is unlikely to undergo significant changes over the QPO cycle. Consequently, we propose that the observed significant phase difference between $T_{\rm in}$ and $F_{\rm disk}$, which causes the $F_{\rm disk}$-$T_{\rm in}$ relation to deviate from the expected behavior of $F_{\rm disk}\tilde{\propto} T^4_{\rm in}$, could be indicative of variations in the accretion geometry over the QPO cycle. In the context of corona/jet precession model, the thin disk is held stationary by viscosity \citep{1975ApJ...195L..65B}. However, the precessing corona and/or jet could obscure different disk azimuths at different precession phases, leading to the variable inner disk emission \citep{2018ApJ...858...82Y,2022MNRAS.511..255N}. Additionally, a precessing corona/jet base could illuminate different azimuthal regions of the disk, thereby heating the disk atmosphere and causing variations in the disk temperature \citep{2016MNRAS.460.2796S,2022MNRAS.511..255N}. Considering quasi-periodic heating of the approaching and receding sides of the disc by a precessing Comptonizing region, one might anticipate a typical phase difference of roughly $0.25$ QPO cycles between the variation in apparent $T_{\rm in}$ (and also the thermal emission) and Comptonized emission reaching the observer directly \citep{2016MNRAS.460.2796S}. This is because the approaching side of the disk is consistently illuminated by the Comptonizing region when the normal of the Comptonizing region aligns perpendicularly to the observer's line of sight \citep[see Figure 12 of][]{2016MNRAS.460.2796S}. However, we would also anticipate that the phase difference should be influenced by the GR effects (e.g. light bending effects), leading to slight deviations from the expected value. Moreover, the situation becomes more complicated when considering the reflection component, where the phase of modulations in apparent temperature and disk flux with respect to the reflected flux would be determined by the azimuthal angle of the observer.
An alternative possibility is the scenario proposed by \citet{2023ApJ...948..116M}, in which the jet and inner disk ring precess together. According to this model, the inner region of the jet and the innermost part of the accretion disk are somehow coupled, leading to a situation where the precession of the disk induces a corresponding precession of the jet. In this case, the peak phases of the variation in the thermal and nonthermal emissions are expected to roughly coincide. As the geometrical alignment between the jet and the inner disk remains constant, the illumination pattern also remains stable throughout the precession cycle. Consequently, the effective temperature ($T_{\rm eff}$) of the accretion disk is expected to remain constant. It is important to note that the fitted peak temperature ($T_{\rm in}$) may be affected by electron scattering and general relativistic (GR) effects, and can be expressed as:
\begin{equation}
    T_{\rm in}=f_{\rm GR}\left(i,a_*\right)f_{\rm col}T_{\rm eff},
\end{equation}
where $f_{\rm GR}$ represents the fractional change of the fitted color temperature due to the GR effects, $i$ is the inclination angle, $a_*$ is the black hole spin, and $f_{\rm col}$ denotes the hardening factor \citep{1997ApJ...482L.155Z}. Numerical simulations suggest a canonical value of $f_{\rm col}\sim1.6-1.8$ \citep{1995ApJ...445..780S,2005ApJ...621..372D,2006ApJ...636L.113S}, whereas $f_{\rm GR}$ strongly depends on the inclination. As demonstrated in \citet{1975ApJ...202..788C} and \citet{1997ApJ...482L.155Z}, the GR effects cause the spectrum to be redshifted at small inclinations, but blueshifted at large inclinations. This indicates that the apparent temperature could display a positive correlation with the inclination. However, for the accretion disk flux over the precession period, it is primarily influenced by the projected area, and is therefore expected to be negatively correlated with the inclination. Consequently, a large phase difference (roughly 0.5 QPO cycles) between the $T_{\rm in}$ variation and the $F_{\rm disk}$ variation could be potentially observed. Furthermore, $f_{\rm col}$ may also vary throughout the precession cycle, due to changes in the optical depth of the electron scattering with the inclination. This effect could also result in a higher color temperature when the disk precesses to an edge-on orientation. Considering the potential presence of a hybrid corona/jet configuration in this source, our findings may suggest a scenario in which the corona, inner disk, and jet are coupled and precess together. It is worth noting that no significant modulations in the polarization degree and polarization angle over the QPO cycle were observed through the phase-resolved polarimetric analysis \citep{2024ApJ...961L..42Z}. If the QPO is produced by the precession of the inner hot flow, findings from the QPO phase-resolved polarimetry suggest high inclinations of the source \citep[$\gtrsim60^\circ$, see][]{2015ApJ...807...53I}. To conduct a more precise phase-resolved analysis for a better comparison with theoretical predictions, obtaining simultaneous polarization, high-statistics timing, and broadband spectroscopy data is crucial. This can be accomplished with the upcoming mission, the enhanced X-ray Timing and Polarimetry Observatory \citep{2019SCPMA..6229502Z}.

In summary, we have performed a detailed phase-resolved analysis of type-C QPOs from the newly discovered BHXRB Swift J1727.8--1613. Our findings indicate that both the source flux and spectral shape exhibit variations over the QPO cycle. In particular, the variability in reflection fraction and disk emission, along with a significant phase difference of the variation in the disk temperature compared to the variation in the disk flux, align with the concept that the QPO variability is not primarily intrinsic to the accretion flow but geometric. 

\begin{acknowledgments}
We are grateful to the anonymous referees for constructive comments that helped us improve this paper. This research has made use of data obtained from the High Energy Astrophysics Science Archive Research Center (HEASARC), provided by NASA’s Goddard Space Flight Center, and the Insight-HXMT mission, a project funded by China National Space Administration (CNSA) and the Chinese Academy of Sciences (CAS). This work is supported by  the National Key R\&D Program of China (2021YFA0718500) and the National Natural Science Foundation of China under grants, 12333007, 12173103, 12027803, U2038101 and U1938103. This work is partially supported by International Partnership Program of Chinese Academy of Sciences (Grant No.113111KYSB20190020). L. D. Kong is grateful for the financial support provided by the Sino-German (CSC-DAAD) Postdoc Scholarship Program (57251553).
P. J. Wang is grateful for the financial support provided by the Sino-German (CSC-DAAD) Postdoc Scholarship Program (57678375).
\end{acknowledgments}

%




\appendix
\section{Hilbert-Huang Transform Method}
\label{appendix1}
\begin{figure*}
\centering
    \includegraphics[width=0.9\textwidth]{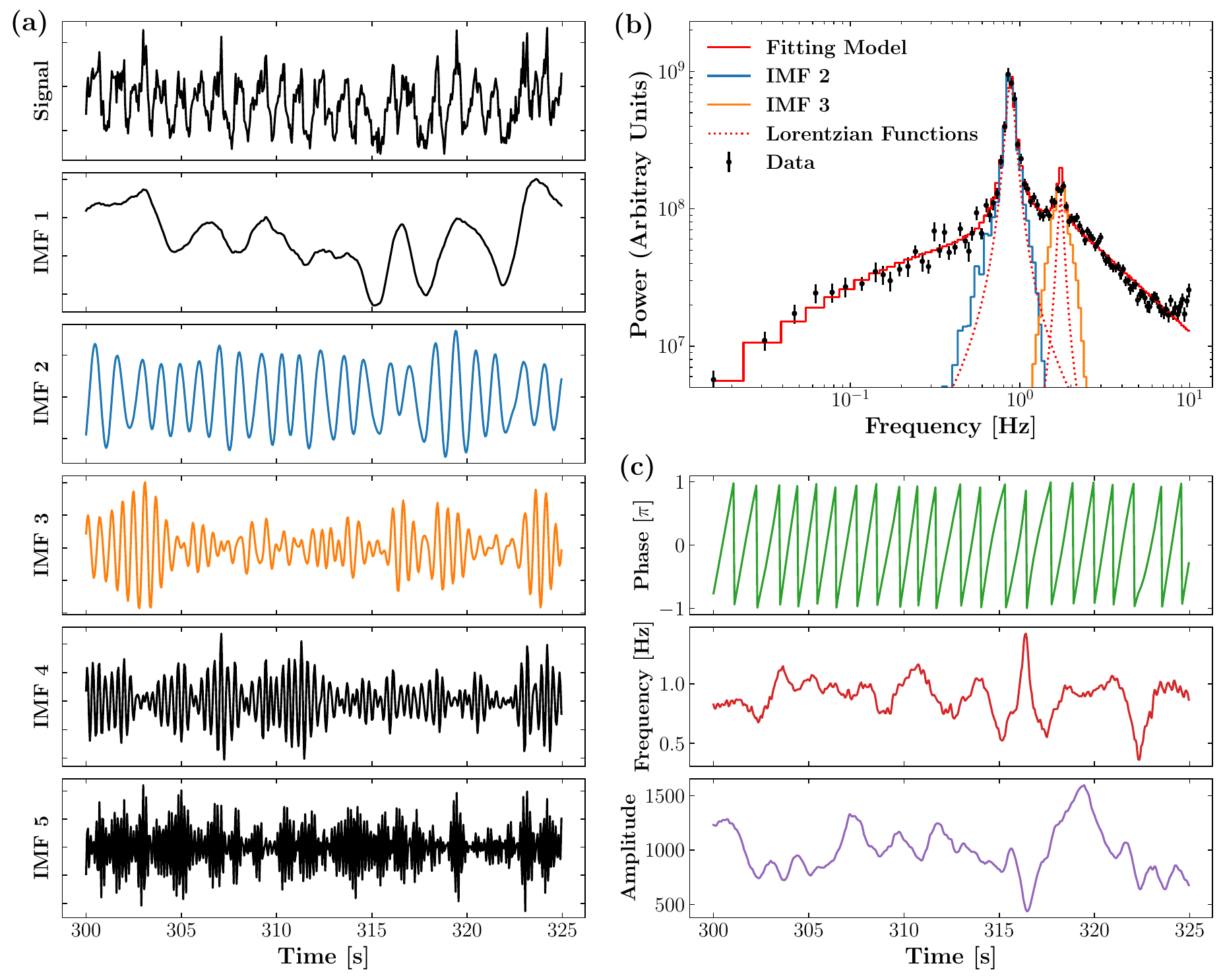}
    \caption{Hilbert–Huang transform analysis of a representative example of a 25 s long light curve with a 0.05 s time resolution of LE in the energy range of 1--10 keV. (a) Representative example of a 25 s long light curve and corresponding five IMFs obtained from VMD algorithm. (b) Power density spectra of the original LE light curve, as well as the second and third IMFs. (c) The corresponding phase function, instantaneous frequency, and amplitude of the QPO (the second) IMF, obtained from the Hilbert transform. In panel (b), we additionally include the fitting model and best-fit Lorentzian functions of the QPO fundamental and second harmonics in the plot.} \label{fig:A1}
\end{figure*}

To obtain the QPO waveform and extract the QPO phase-resolved spectra, we perform the HHT analysis on the NICER and Insight-HXMT data. Figure~\ref{fig:A1} presents a representative HHT analysis of Insight-HXMT data. For a comprehensive review of the method, see \citet{2023ApJ...957...84S,2024ApJ...965L...7S}.

\section{MCMC Parameter Probability Distributions}
\label{appendix2}
\begin{figure*}
\centering
    \includegraphics[width=0.9\textwidth]{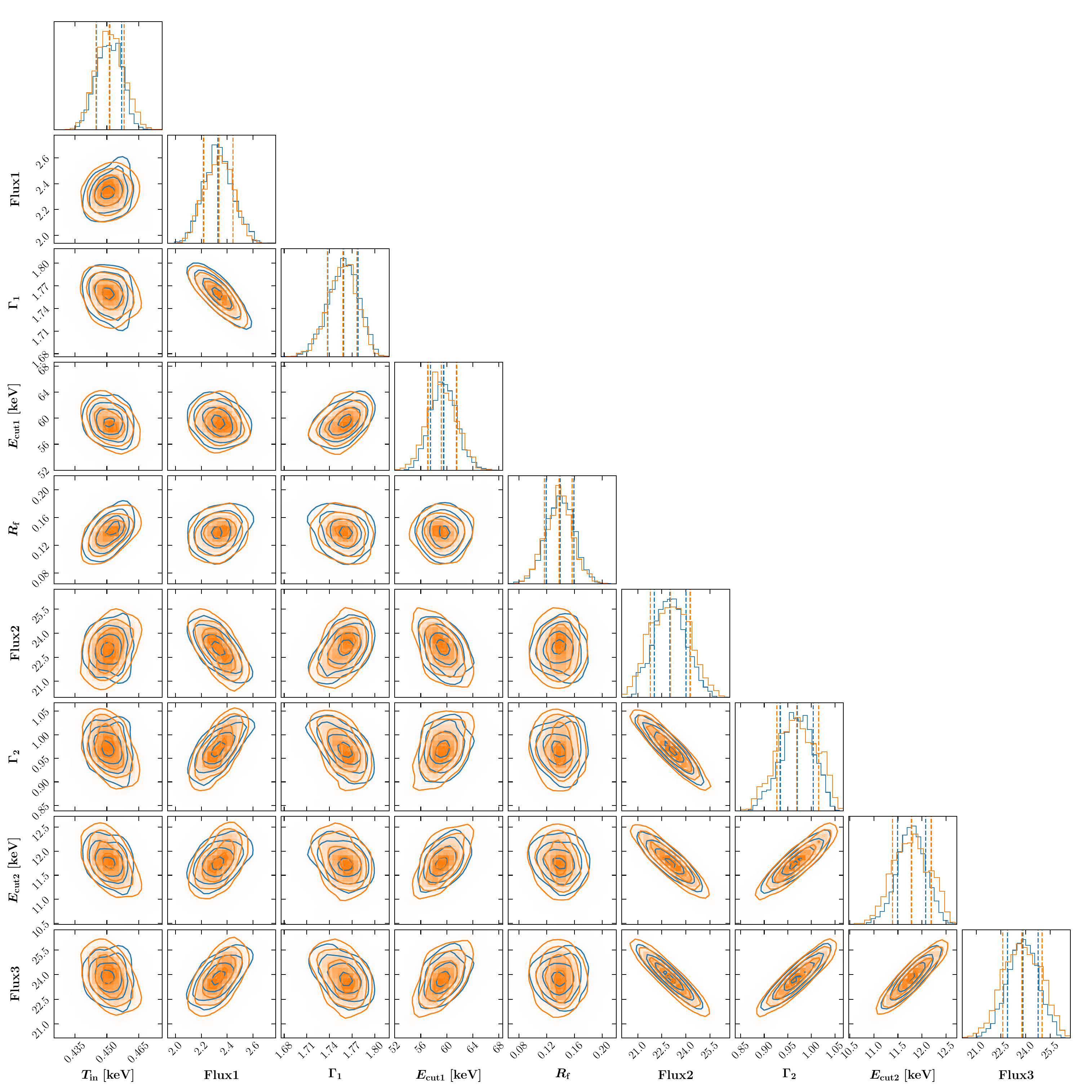}
    \caption{One- and two-dimensional projections of the posterior probability distributions, and the 0.16, 0.5, and 0.84 quantile contours derived from the MCMC analysis for each free spectral parameter, from the joint spectral fitting of NICER and Insight-HXMT data. To test the convergence, we compare the one- and two-dimensional projections of the posterior distributions from the first (blue) and second (orange) halves of the chain. This illustration corresponds to the spectral fitting of the QPO peak phase.} \label{fig:A2}
\end{figure*}

This appendix contains corner plots of spectral parameters from an example MCMC analysis. We use the Goodman–Weare algorithm with 32 walkers and a total length of 50,000 to perform the MCMC analysis, and the initial 2000 elements are discarded as the burn-in period during which the chain reaches its stationary state. In Figure~\ref{fig:A2}, we compare the one- and two-dimensional projections of the posterior distributions for each parameter from the first and second halves of the chain to test the convergence. The contour maps and probability distributions are plotted using the corner package \citep{2016JOSS....1...24F}.

\bibliography{sample631}{}
\bibliographystyle{aasjournal}



\end{document}